\documentclass[sigconf]{acmart}
 \usepackage{textcomp, libertine}
\usepackage{booktabs} 
\usepackage[normalem]{ulem}
\usepackage{multirow}
\usepackage{bm}
\useunder{\uline}{\ul}{}

\usepackage{subcaption}
\usepackage[para]{footmisc}
\usepackage{enumitem}
\usepackage[ruled]{algorithm2e}
\usepackage{amssymb}
\usepackage{dsfont}
\usepackage{pifont}
\newcommand{\cmark}{\ding{51}}%
\newcommand{\xmark}{\ding{55}}%
\hypersetup{
,bookmarksnumbered=true%
,hypertexnames=false%
,breaklinks=true%
,colorlinks=true%
,linkcolor=black%
,citecolor=black%
,urlcolor=black%
}

\definecolor{tealblue}{rgb}{0.21, 0.46, 0.53}
\definecolor{wildstrawberry}{rgb}{1.0, 0.26, 0.64}
\definecolor{ao(english)}{rgb}{0.0, 0.5, 0.0}

\def\ignore#1{}

\copyrightyear{2020}
\acmYear{2020}
\setcopyright{acmcopyright}\acmConference[SIGIR '20]{Proceedings of the 43rd International ACM SIGIR Conference on Research and Development in Information Retrieval}{July 25--30, 2020}{Virtual Event, China}
\acmBooktitle{Proceedings of the 43rd International ACM SIGIR Conference on Research and Development in Information Retrieval (SIGIR '20), July 25--30, 2020, Virtual Event, China}
\acmPrice{15.00}
\acmDOI{10.1145/3397271.3401110}
\acmISBN{978-1-4503-8016-4/20/07}

\begin{document}

\settopmatter{printacmref=false, printfolios=false}
\fancyhead{}

\title{Open-Retrieval Conversational Question Answering}

\author{Chen Qu$^1$ \quad  Liu Yang$^1$ \quad Cen Chen$^2$ \quad Minghui Qiu$^3$ \quad  W. Bruce Croft$^1$  \quad Mohit Iyyer$^1$}

\affiliation{%
	\institution{
		$^1$ University of Massachusetts Amherst \quad
		$^2$ Ant Financial \quad
		$^3$ Alibaba Group } 
}
\email{{chenqu, lyang, croft, miyyer}@cs.umass.edu,
chencen.cc@antfin.com, 
minghui.qmh@alibaba-inc.com}

\begin{abstract}
Conversational search is one of the ultimate goals of information retrieval. Recent research approaches conversational search by simplified settings of response ranking and conversational question answering, where an answer is either selected from a \textit{given} candidate set or extracted from a \textit{given} passage. These simplifications neglect the fundamental role of retrieval in conversational search. To address this limitation, we introduce an open-retrieval conversational question answering (ORConvQA) setting, where we \textit{learn to retrieve evidence from a large collection} before extracting answers, as a further step towards building functional conversational search systems. We create a dataset, OR-QuAC, to facilitate research on ORConvQA. We build an end-to-end system for ORConvQA, featuring a retriever, a reranker, and a reader that are all based on Transformers. Our extensive experiments on OR-QuAC demonstrate that a learnable retriever is crucial for ORConvQA. We further show that our system can make a substantial improvement when we enable history modeling in all system components. Moreover, we show that the reranker component contributes to the model performance by providing a regularization effect. Finally, further in-depth analyses are performed to provide new insights into ORConvQA.
\end{abstract}


\maketitle

{\fontsize{8pt}{8pt} \selectfont
\textbf{ACM Reference Format:}\\
Chen Qu, Liu Yang, Cen Chen, Minghui Qiu, W. Bruce Croft, and Mohit Iyyer. 2020. Open-Retrieval Conversational Question Answering. In \textit{Proceedings of the 43rd International ACM SIGIR Conference on Research and Development in Information Retrieval (SIGIR '20), July 25--30, 2020, Virtual Event, China.} ACM, New York, NY, USA, 10 pages. \url{https://doi.org/10.1145/3397271.3401110}}

\section{Introduction}
\label{sec:intro}

Conversational search is an embodiment of an iterative and interactive information retrieval (IR) system that has been studied for decades~\cite{Belkin1994CasesS,i3r,Oddy1977Information}.
Due to the recent rise of intelligent assistant systems, such as Siri, Alexa, AliMe, Cortana, and Google Assistant, a growing part of the population is moving their information-seeking activities to voice or text based conversational interfaces. This trend is closely related to the revival of research interest in question answering (QA) and conversational QA (ConvQA). ConvQA can be considered as a simplified setting of conversational search~\cite{hae}. A significant limitation of this setting is that an answer is either extracted from a \textit{given} passage~\cite{ham} or selected from a \textit{given} candidate set~\cite{Yang2018ResponseRW}. This simplification neglects the fundamental role of retrieval in conversational search. To address this issue, we introduce an open-retrieval ConvQA (ORConvQA) setting, where we \textit{learn to retrieve evidence from a large collection} before extracting answers.

We illustrate the importance of ORConvQA by characterizing the task and discussing the considerations of an ORConvQA dataset as follows. A comparison between ORConvQA and related tasks is presented in Table~\ref{tab:intro}. 

\begin{table}[t]
\caption{Comparison of selected QA tasks on the dimensions of open-retrieval (OR), conversational (Conv), information-seeking (IS), and whether motivated by genuine information needs (GIN). The symbol ``-'' suggests a mixed situation.}
\label{tab:intro}
\vspace{-0.3cm}
\begin{tabular}{@{}lllll@{}}
\toprule
Task \& Example Data                                   & OR      & Conv      & IS      & GIN \\ \midrule
Open-Retrieval QA \cite{orqa,Das2019MultistepRI}       & \cmark  & \xmark    & \textbf{-} & \textbf{-}   \\
Response Ranking w/ UDC \cite{udc,Yang2018ResponseRW,smn}  & \xmark  & \cmark    & \cmark  & \cmark   \\
Conversational MC w/ CoQA \cite{sdnet,graphflow}       & \xmark  & \cmark    & \xmark  & \xmark   \\
Conversational MC w/ QuAC \cite{ham,flowqa}            & \xmark  & \cmark    & \cmark  & \cmark   \\
ORConvQA w/ OR-QuAC (this work)                        & \cmark  & \cmark    & \cmark  & \cmark   \\ \bottomrule
\end{tabular}
\end{table}

1. \textbf{Open-retrieval}. This is a defining property of ORConvQA. 
In recent ConvQA datasets~\cite{quac,coqa}, the ConvQA task is formulated as a conversational machine comprehension (MC) problem with the goal being to extract or generate an answer from a given gold passage. 
This setting can be impractical in real-world applications since the gold passage is not always available, or there could be no ground truth answer in the given passage. Instead of being given the passage, a ConvQA system should be able to retrieve candidate passages from a collection. In particular, it is desirable if this retriever is learnable and can be fine-tuned on the downstream ConvQA task, instead of adopting fixed heuristic retrieval functions like TF-IDF or BM25. Moreover, the retrieval process should be open in terms of retrieving from a large collection instead of reranking a small number of passages in a closed set.

2. \textbf{Conversational}. Being conversational reflects the interactive nature of a search activity. The important problem of user interaction modeling in IR can be formulated as conversation history modeling in this scenario. 

3. \textbf{Information-seeking}. 
An information-seeking conversation typically requires multiple turns of information exchange to allow the seeker to clarify an information need, provide feedback, and ask follow up questions. 
In this process, answers are revealed to the seeker through a sequence of interactions between the seeker and the provider. These answers are generally longer than the entity-based answers in factoid QA. 

4. \textbf{Genuine information needs}. An information-seeking conversation is closer to real-world scenarios if the seeker is genuinely seeking an answer. In SQuAD~\cite{squad} and CoQA~\cite{coqa}, the seekers' information needs are not genuine because they have access to the passage and thus have the target answer in mind when asking the question. These questions are referred to as ``back-written questions''~\cite{reqa} and have been reported to have more lexical overlap with their answers in SQuAD~\cite{reqa}. This undesirable property makes the models learned from such datasets less practical.

To the best of our knowledge, there has not been a publicly available dataset that satisfies all the properties we discussed as shown in Table~\ref{tab:intro}. We address this issue by aggregating existing data to create the \textit{OR-QuAC} dataset. The QuAC~\cite{quac} dataset offers information-seeking conversations that are collected with no seekers' prior knowledge of the passages. We extend QuAC to an open-retrieval setting by creating a collection of over 11 million passages using the whole Wikipedia corpus. Another important resource used in our aggregation process is the CANARD dataset~\cite{canard} that offers context-independent rewrites of QuAC questions. Some initial questions in QuAC conversations are underspecified. This makes conversation difficult to interpret in an open-retrieval setting. We make these dialogs self-contained by replacing the \textit{initial} question in a conversation with its rewrite from CANARD. Our data has 5,644 dialogs with 40,527 questions. We release OR-QuAC to the community to facilitate research on ORConvQA.

In addition to proposing ORConvQA and creating the OR-QuAC dataset, we develop a system for ORConvQA following previous work on open-retrieval QA~\cite{orqa}. Our end-to-end system features a retriever, a reranker, and a reader that are all based on Transformers~\cite{transformer}. We enable history modeling in all components by concatenating history questions to the current question.
The passage retriever first retrieves the top $K$ relevant passages from the collection given a question and its history. The reranker and reader then rerank and read the top passages to produce an answer. The training process contains two phases, a pretraining phase for the retriever and a concurrent learning phase for all system components.

Specifically, our retriever adopts a dual-encoder architecture~\cite{reqa, orqa,Das2019MultistepRI,dpr} that uses separate ALBERT~\cite{albert} encoders for questions and passages. The question encoder also encodes conversation history. 
After being pretrained, the passage encoder is frozen and encodes all passages in the collection offline.
The reranker and the reader share the same BERT~\cite{bert} encoder. It encodes the input sequence of a concatenation of the question, history, and each relevant passage to contextualized representations for reranking and answer extraction. We incorporate shared-normalization~\cite{shared-norm} in our system to enable comparison among the candidate passages. In the concurrent learning phase, we encode the question and the history to dense vectors with the question encoder for an efficient retrieval with maximum inner product search (MIPS)~\cite{mips,faiss}. The top retrieved passages are fed to the reranker and reader for a concurrent learning of all model components. 

We conduct extensive experiments on our OR-QuAC dataset. First, we show that our system without any history information has comparable performance with a conversational version of BERTserini~\cite{bertserini} that considers history. 
This improvement demonstrates the importance of a learnable retriever in ORConvQA. We further show that our system can make a substantial improvement when we enable history modeling in all system components. Moreover, we conduct in-depth analyses on model ablation and configuration to provide insights for the ORConvQA task. We show that our reranker component contributes to the model performance by providing a regularization effect. We also demonstrate that the initial question of each dialog is crucial for our system to understand the user's information need. Our code and data are available for research purposes.\footnote{\url{https://github.com/prdwb/orconvqa-release}}

\section{Related Work}
\label{sec:relatedwork}
Our work is closely related to several research topics, including QA, open domain QA, ConvQA, and conversational search. We mainly discuss retrieval based methods since they tend to offer more informative responses~\cite{Yang2018ResponseRW} and thus better fit for information-seeking tasks than generation based methods.

\textbf{Question Answering}. 
One of the first modern reformulations of the QA task dates back to the TREC-8 Question Answering Track~\cite{trec8}. Its goal is to answer 200 fact-based, short-answer questions by leveraging a large collection of documents. A retrieval module is crucial in this task to retrieve relevant passages for answer extraction. As an increasing number of researchers in the natural language processing (NLP) community moving their focus to answer extraction and generation methods, the role of retrieval has been gradually overlooked. As a result, many popular QA tasks and datasets either follow an answer selection setting \cite{wikiqa,trecqa,tanda} or a machine comprehension setting~\cite{squad,squad2,GoogleNQ,newsqa}. 
In real-world scenarios, it is less practical to assume we are given a small set of candidate answers or a gold passage. Therefore, in this work, we make the retrieval component as one of our focuses in the task formulation and model architecture.

\textbf{Open Domain Question Answering}.
In contrast to the tasks that offer a pre-selected passage for answer extraction, open domain QA tasks provide the model with access to a large corpus~\cite{quasar} or at least a set of candidate documents for each question~\cite{Marco,TriviaQA,searchqa,quasar,WikiPassageQA}. 
DrQA~\cite{drqa} and BERTserini~\cite{bertserini} present an end-to-end open domain QA system by using a TF-IDF/BM25 retriever and a neural reader. 
Some previous work~\cite{rank_paragraph,htut_rank,adaptive_doc_retrieval,r3} learns to rerank or select from a closed set of passages for open domain QA. These methods may not scale well to an open-retrieval setting. 
Recently, \citet{orqa}, \citet{Das2019MultistepRI}, and \citet{dpr} adopt a dual-encoder architecture to construct a learnable retriever and demonstrate their methods are scalable to large collections. ReQA~\cite{reqa} also uses a similar architecture to retrieve sentence-level answers directly. 
Although these works are limited to single turn QAs, they are valuable resources for us to study how to extend ConvQA to an open-retrieval setting.

\textbf{Conversational Question Answering}.
Similar to the answer selection and MC tasks in single-turn QAs, existing ConvQA research is mostly limited to response ranking~\cite{Yang2018ResponseRW,udc,Yan2016ShallIB,Yan2016LearningTR,Tao2019MultiRepresentationFN,smn,hybrid,iart} and conversational MC~\cite{quac,coqa,hae,ham,flowqa,sdnet,bertflowdelta,graphflow}, where the role of retrieval is also neglected. Open-retrieval is particularly important to ConvQA since the answers of the questions from the same dialog may not necessarily come from the same passage. The model needs to learn to retrieve passages for each dialog question. Another challenge is to investigate how to enable history modeling not only in the reader but also in the retriever. Moreover, there are no existing datasets that are suitable to study ORConvQA. Therefore, we tackle these research questions in this work.

\textbf{Conversational Search}.
While the concept of conversational search can be traced back to research~\cite{Belkin1994CasesS,i3r,Oddy1977Information} from decades ago, recent years have witnessed its revival. In addition to the ConvQA work mentioned above, researchers are also actively working on other conversation tasks, including conversational recommendation and product search~\cite{Bi2019ConversationalPS,Zhang2018TowardsCS}, user intent prediction~\cite{UserIntentPred}, and question retrieval~\cite{Yang2017NeuralMM}. Another rich body of work targets the user-oriented aspect~\cite{Qu2018AnalyzingAC,Chuklin2018ProsodyMF,Trippas2018InformingTD,misc,answer_interaction,Trippas2017HowDP,Trippas2019TowardsAM} for conversational information seeking. Our work extends ConvQA to an open-retrieval setting as another fundamental step towards conversational search.

\section{The OR-QuAC Dataset}
\label{sec:dataset}
\begin{figure}[t]
    \centering
    \includegraphics[width=0.475\textwidth]{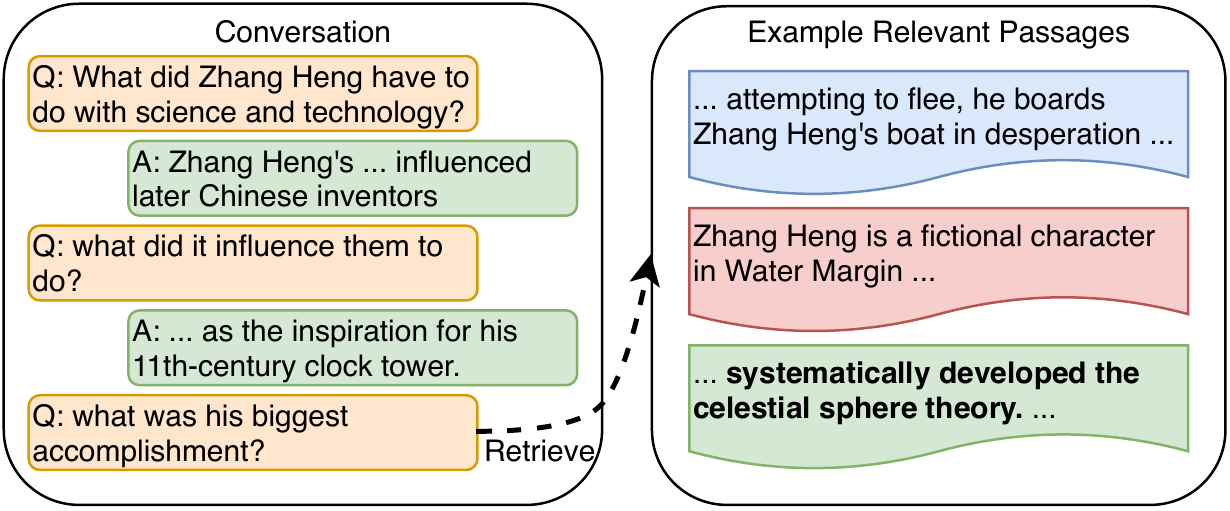}
    \vspace{-0.5cm}
    \caption{A partial OR-QuAC dialog and example relevant passages retrieved from the collection by TF-IDF.}
    \label{fig:example}
\end{figure}

The OR-QuAC dataset enhances QuAC by adapting it to an open-retrieval setting. It is an aggregation of three existing datasets: (1) the QuAC dataset~\cite{quac} that offers information-seeking conversations, (2) the CANARD dataset~\cite{canard} that consists of context-independent rewrites of QuAC questions, and (3) the Wikipedia corpus that serves as the knowledge source of answering questions. An example of OR-QuAC is presented in Figure~\ref{fig:example}. We will describe the data construction process in the following sections. 

\subsection{\mbox{Self-Contained Information-seeking Dialogs}}
\label{subsec:data-dialogs}
The QuAC (Question Answering in Context) dataset~\cite{quac} is designed for modeling information-seeking conversations. It consists of real human-human dialogs between an information seeker and an information provider. The seeker tries to learn about a hidden Wikipedia text by asking a sequence of freeform questions. She/he only has access to the title and a summary of the article. This simulates a genuine information need. The provider answers each question by indicating a short span of the given passage. This dataset poses unique challenges because its questions are more open-ended, sometimes unanswerable, or only meaningful within the dialog context~\cite{quac}.

A drawback of QuAC is that many dialogs are not self-contained. This is typically caused by incomplete initial questions. A QuAC dialog is motivated by a general and underlying information need. During the data collection process, this information need is provided to both the seeker and provider before initiating the dialog. Therefore, the seeker might not necessarily reiterate this information need when asking the first question. For example, a seeker in QuAC is instructed to learn about \textit{Zhang Heng}, a Chinese polymathic scientist. The very first question the seeker asked was "\textit{what did he have to do with science and technology?}". Such underspecified and ambiguous initial questions become an issue in the open-retrieval setting because they make the conversation difficult to interpret.

We tackle this issue by replacing \textit{initial} questions in QuAC with their context-independent \textit{rewrites} provided by the CANARD dataset. For example, the rewrite for the previously mentioned question is "\textit{What did Zhang Heng have to do with science and technology?}". We do the replacement for the first questions only. This makes a dialog self-contained while keeping the history dependencies within the dialog untouched.

CANARD covers about half of the released QuAC questions. Since the QuAC test set is not publicly available, they use QuAC's development set as their test set and 10\% of QuAC's training set as their development set~\cite{canard}. We follow the data split of CANARD. QuAC questions that not in CANARD are discarded. The data statistics of our derived dataset, OR-QuAC, are presented in Table~\ref{tab:data-stat}.

\begin{table}[t]
\caption{Data statistics of the OR-QuAC dataset.}
\label{tab:data-stat}
\vspace{-0.4cm}
\begin{tabular}{@{}llll@{}}
\toprule
Items                                                                                    & Train      & Dev        & Test       \\ \midrule
\# Dialogs                                                                               & 4,383      & 490        & 771        \\
\# Questions                                                                             & 31,526     & 3,430      & 5,571      \\
\# Avg. Question Tokens                                                                  & 6.7        & 6.6        & 6.7        \\
\# Avg. Answer Tokens                                                                    & 12.5       & 12.6       & 12.2       \\
\# Avg. Questions / Dialog                                                               & 7.2        & 7.0        & 7.2        \\
\begin{tabular}[c]{@{}l@{}}\# Min/Avg/Med/Max \\ History Turns / Question\end{tabular} & 0/3.4/3/11 & 0/3.3/3/11 & 0/3.4/3/11 \\ \bottomrule
\end{tabular}
\end{table}

\subsection{Collection}
\label{subsec:data-collection}
We use the whole Wikipedia corpus to construct a collection since passages in QuAC are from Wikipedia. We use the English Wikipedia dump from 10/20/2019.\footnote{https://dumps.wikimedia.org/enwiki/20191020/} The Wikipedia passages in QuAC were downloaded via PetScan\footnote{http://petscan.wmflabs.org/}~\cite{quac}, and thus, the exact date for the data dump is unavailable. Therefore, we use the latest data dump instead of trying to match the date of QuAC. We then use the WikiExtractor\footnote{https://github.com/attardi/wikiextractor} to extract and clean text from the data dump, resulting in over 5.9 million Wikipedia articles. After this, we split the articles into chunks with at most 384 wordpieces using the tokenizer of BERT, following \citet{orqa}. The split is done greedily while preserving sentence boundaries.
These chunks are referred to as \textit{passages}. Less than 0.5\% of known answers are split into different passages. Their corresponding questions are considered as unanswerable during training. We do the split to make the passages fit for Transformer based retrievers and readers. Moreover, \citet{bertserini} reported that the paragraph level is the best granularity for an end-to-end retrieve-and-read framework compared to the article and sentence levels. 
They believe the reason is that an article may contain non-relevant content that distracts the reader while a sentence may lack context information. For an open-retrieval setting, we prefer passage-level retrieval over article-level since a full article would be harder to represent with a fixed-length dense vector. 

Since the paragraphs in QuAC may not be exactly the same as those in the Wikipedia dump given the difference in the dates of the dumps, we conduct the same split process for QuAC paragraphs and replace the Wikipedia passages with QuAC passages that have the same article titles.
The positions of the ground truth answer spans are mapped to the new passages. The resulting collection has over 11 million passages for retrieval.

Due to the synthetic nature of this dataset, the answers of the questions in the same dialog are distributed in the same section of text. In real world, questions and answers in a dialog may be distributed at different locations of the corpus. This is a limitation of our dataset.

\section{An End-to-end ORConvQA System}
\label{sec:our-approach}
\begin{figure*}[th]
    \centering
    \includegraphics[width=0.9\textwidth]{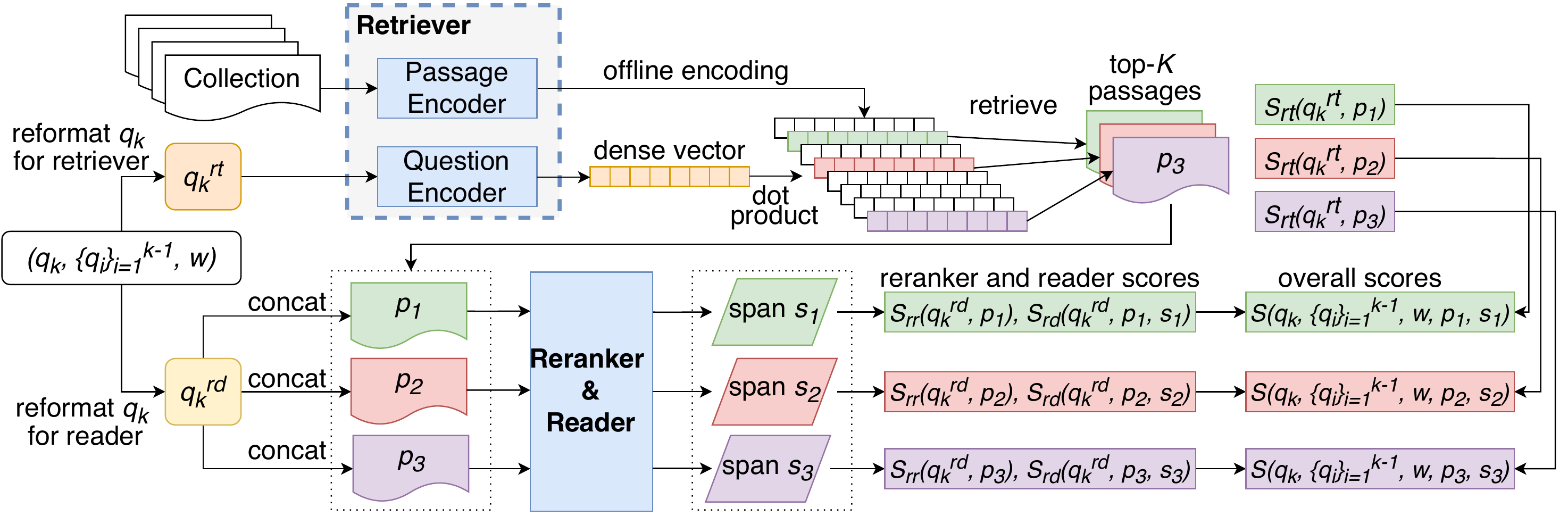}
    \vspace{-0.3cm}
    \caption{Architecture of our end-to-end ORConvQA model. The input is the current question $q_k$, all history questions $\{q_i\}_{i=1}^{k-1}$, and a history window size $w$. The retriever first retrieves top-$K$ relevant passages from the collection and generates retriever scores $S_{rt}$. The reranker and reader then rerank and read the top passages to produce an answer span for each passage and generate reranker and reader scores, $S_{rr}$ and $S_{rd}$. The system outputs the answer span with the highest overall score $S$.
    }
    \label{fig:model}
\end{figure*} 

In this section, we first formally define the task of open-retrieval conversational QA. We then describe our end-to-end system that deals with this task and explain the intuitions behind it.

\subsection{Task Definition}
\label{subsec:task}
The ORConvQA task is defined as follows. Given the $k$-th question $q_k$ in a conversation, and all history questions $\{q_i\}_{i=1}^{k-1}$ preceding $q_k$, the task is to predict an answer $a_k$ for $q_k$ using a passage collection $C$. In an extractive setting, $a_k$ is a text span of a passage in $C$. We do not assume we have access to ground truth history answers since it is impractical in real-world scenarios.  

Extractive models are trained on the supervision signals of the position of a span in the gold passage. Previous works~\cite{drqa,orqa,Das2019MultistepRI} present a distantly-supervised setting, where they only have access to QA pairs without gold passages. This setting heavily relies on a heuristic that a positive passage should contain an exact match of the known answer. Short and entity-based answers can often be discovered in multiple passages, meaning that positive passages are highly substitutable. In information-seeking conversations that are motivated by genuine information needs, however, the answers are typically much longer. For example, QuAC answers have 12 tokens on average while SQuAD and CoQA answers have 3~\cite{coqa}. It is common that the retrieved passages do not contain exact matches of the known answers, making many training examples useless. 
To tackle this, we adopt a fully-supervised setting: we assume we have access to gold passages so that we can include them if they are not present in the retrieval results and use the ground-truth answer spans. This is done at \textit{training} time only. Although this is a limitation, it does not conflict with the learnable retriever we promote. We will work on a weak supervision method that is suitable for information-seeking conversations in our future work.

\subsection{Model Overview}
\label{subsec:model-overview}
We now present an end-to-end system that deals with the ORConvQA task described in Section~\ref{subsec:task}. Our system consists of three major components, a passage retriever, a passage reranker, and a passage reader. The reranker and reader are based on the same encoder. All components are learnable. As described in Figure~\ref{fig:model}, the passage retriever first retrieves top-$K$ relevant passages from the collection given a question and its history. The passage reranker and reader then rerank and read the top passages to produce an answer. History modeling is enabled in all components. We will describe each component in detail in the following sections.

\subsection{Passage Retriever}
\label{subsec:model-retriever}
We present the retriever module in the upper-left part of Figure~\ref{fig:model}. We follow previous research~\cite{reqa,orqa,Das2019MultistepRI} by using a dual-encoder architecture to construct a learnable retriever. This architecture features separated encoders for questions and passages. The retriever score is then defined as the dot product of the hidden representations of a question and a passage. We use two ALBERT~\cite{albert} models for both encoders. ALBERT is a lite BERT~\cite{bert} model for learning bidirectional language representations from Transformers~\cite{transformer}. It reduces the parameters of BERT by cross-layer parameter sharing and embedding parameters factorization~\cite{albert}.

Given all available history questions $\{q_i\}_{i=1}^{k-1}$, we first identify those that are in a history window with the size $w$. These questions are denoted as $\{q_i\}_{i=k-w}^{k-1}$. We then construct a concatenation of $\{q_i\}_{i=k-w}^{k-1}$ and $q_k$. We prepend the initial question $q_1$ of the conversation to the concatenation if $q_1$ is not already included. The initial question $q_1$ typically contains an information need that is pertinent to the entire conversation as explained in Section~\ref{subsec:data-dialogs}. The reformatted question
for the retriever is denoted as $q_k^{rt}$. For an ALBERT based question encoder, the input sequence would be ``\texttt{[CLS]} $q_1$ \texttt{[SEP]} $q_{k-w}$ \texttt{[SEP]} $\cdots$ \texttt{[SEP]} $q_{k-1}$\texttt{[SEP]} $q_k$ \texttt{[SEP]}''. All questions are in the same segment. \texttt{[CLS]} and \texttt{[SEP]} are special tokens introduced in BERT~\cite{bert}. We then take the \texttt{[CLS]} representation and project it to a 128-dimensional vector as the question representation following \citet{orqa}. Formally,
\begin{equation}\label{eqn:question-encoder}
\footnotesize
\begin{aligned}
v_q = W_q\ \text{ALBERT}_q (q_k^{rt}) \texttt{[CLS]}
\end{aligned}
\end{equation}
where $\text{ALBERT}_q$ is the question encoder, $W_q$ is the projection matrix for the question \texttt{[CLS]} representation, and $v_q \in \mathbb{R}^{1 \times 128}$ is the final question representation enhanced with history information. We then follow the same scheme to obtain the passage representation for a passage $p_j$:
\begin{equation}\label{eqn:passage-encoder}
\footnotesize
\begin{aligned}
v_p = W_p\ \text{ALBERT}_p (p_j) \texttt{[CLS]}
\end{aligned}
\end{equation}
where $p_j$ is a passage in the collection, $\text{ALBERT}_p$ is the passage encoder, $W_p$ is the projection matrix for the passage \texttt{[CLS]} representation, and $v_p\in \mathbb{R}^{1 \times 128}$ is the final passage representation. Finally, the retrieval score is computed as 
\begin{equation}\label{eqn:retriever-score}
\footnotesize
\begin{aligned}
S_{rt}(q_k^{rt}, p_j) = v_q v_p^\top 
\end{aligned}
\end{equation}

\subsection{Passage Reader/Reranker}
\label{subsec:model-reader}
Given the current question $q_k$, history questions $\{q_i\}_{i=1}^{k-1}$, the history window size $w$, and one of the retrieved passages $p_j$, the passage reader predicts an answer span within the passage. In contrast to \citet{orqa} and \citet{bertserini}, we introduce reranking into this process with little additional cost. Our reader mostly follows the standard architecture of a BERT based MC model~\cite{bert}. We enhance this model by applying the shared-normalization mechanism proposed by \citet{shared-norm} to enable comparison across all retrieved passages for a question. Similar mechanisms are also adopted by \citet{bertserini} and \citet{orqa}. 

\subsubsection{\textbf{Encoder}}
The reader and reranker share the same BERT encoder. Similar to the retriever, we first construct a reformatted question by concatenating history questions within a history window and the current question. We do not additionally prepend the initial question because the conversation is considered to be grounded to $p_j$. The reformatted question for the reader is denoted as $q_k^{rd}$. We then concatenate a retrieved passage to form the input sequence for the BERT model. Specifically, the input sequence $(q_k^{rd},\ p_j)$ is ``\texttt{[CLS]} $q_{k-w}$ \texttt{[SEP]} $\cdots$ \texttt{[SEP]} $q_{k-1}$\texttt{[SEP]} $q_k$ \texttt{[SEP]} $p_j$ \texttt{[SEP]}'', with $q_k^{rd}$ and $p_j$ in different segments. The BERT model then generates contextualized representations for all tokens in the input sequence:
\begin{equation}\label{eqn:reader-encoder-m}
\footnotesize
\begin{aligned}
v_\texttt{[m]} = \text{BERT} ((q_k^{rd},\ p_j) )\texttt{[m]}
\end{aligned}
\end{equation}
where $v_\texttt{[m]}$ is the representation for the $m$-th token in the input sequence. We also need the sequence representation obtained by
\begin{equation}\label{eqn:reader-encoder-cls}
\footnotesize
\begin{aligned}
v_\texttt{[CLS]} = W_\texttt{[CLS]}\ \text{BERT}((q_k^{rd},\ p_j)) \texttt{[CLS]}
\end{aligned}
\end{equation}
where $W_\texttt{[CLS]}$ is a projection for the \texttt{[CLS]} representation to obtain the sequence representation $v_{\texttt{[CLS]}}$ following \citet{bert}. 

\subsubsection{\textbf{Reranker}}
As shown in Figure~\ref{fig:model}. The reranker components conduct a listwise reranking of the top retrieved passages. The reranking task provides more supervision signals to fine-tune the BERT encoder. The representation learning of the encoder also benefits from a regularization effect for optimizing for multiple tasks. Moreover, the reranking task adds little additional cost to the training process because representations for all tokens, including the \texttt{[CLS]} token, are generated with vectorization in a Transformer architecture.  Specifically, we learn a reranking vector $\mathbf{W}_{rr}$ to project the sequence representation $v_\texttt{[CLS]}$ to a reranking score $S_{rr}$:
\begin{equation}\label{eqn:reranking-score}
\footnotesize
\begin{aligned}
S_{rr}(q_k^{rd}, p_j) = \mathbf{W}_{rr} v_\texttt{[CLS]} 
\end{aligned}
\end{equation}

\subsubsection{\textbf{Reader}}
The reader predicts an answer span by computing scores of each token being the start token and the end token. We learn two sets of parameters, a start vector $\mathbf{W}_{s}$ and an end vector $\mathbf{W}_{e}$, to project token representations to start and end scores:
\begin{equation}\label{eqn:start-end-score}
\footnotesize
\begin{aligned}
S_s(q_k^{rd}, p_j, \texttt{[m]}) = \mathbf{W}_{s} v_\texttt{[m]} \quad, \quad S_{e}(q_k^{rd}, p_j, \texttt{[m]}) = \mathbf{W}_{e} v_\texttt{[m]}
\end{aligned}
\end{equation}
where  $S_s(q_k^{rd}, p_j, \texttt{[m]})$ and $S_e(q_k^{rd}, p_j, \texttt{[m]})$ are the scores for the $m$-th token being the start and end tokens of the answer span. The reader score and overall score will be computed at inference time in Section~\ref{subsec:model-inference}.

\subsection{Training}
\label{subsec:model-training}
Our training procedure contains two phases. The first is the retriever pretraining phase, followed by the concurrent learning phase of the retriever (question encoder), reranker, and reader.

\subsubsection{\textbf{Retriever Pretraining}}
\begin{figure}[t]
    \centering
    \includegraphics[width=0.475\textwidth]{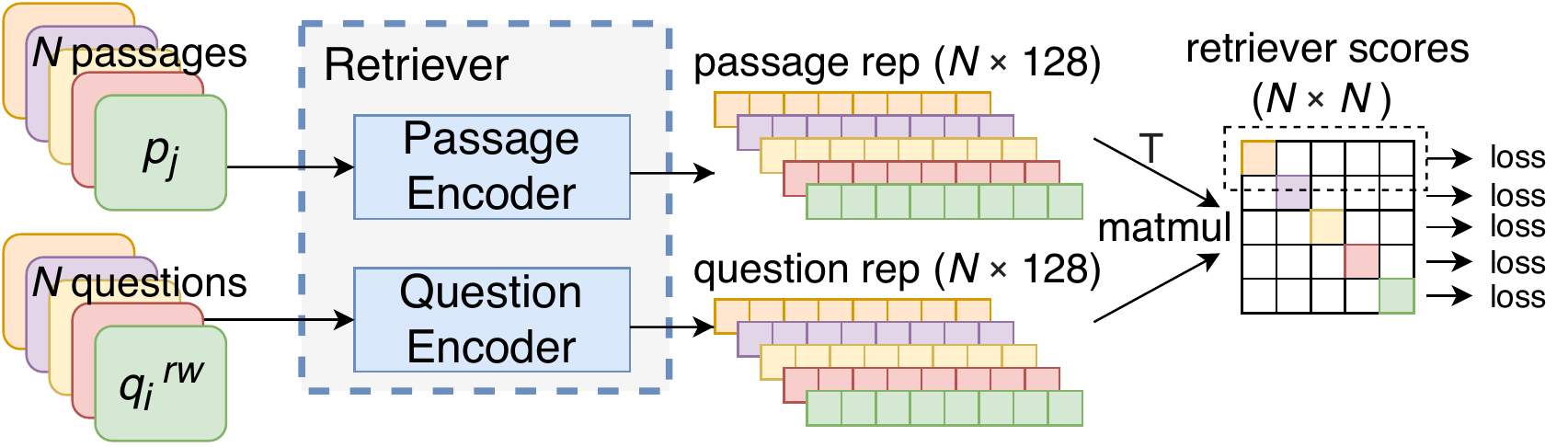}
    \vspace{-0.5cm}
    \caption{Retriever pretraining.}
    \label{fig:retriever}
\end{figure}
We follow previous work~\cite{orqa} to pretrain the retriever so that it gives a reasonable performance in the concurrent learning phase.

In Section~\ref{subsec:model-retriever}, we mentioned that history modeling is enabled in the retriever by prepending history questions. The history window size $w$ is a hyper-parameter and is tunable. In the pretraining phase, however, we would like to train a uniform retriever for every single history window size. Therefore, we use the rewrite in CANARD $q_i^{rw}$ as the reformatted question for a question $q_i$ in the pretraining phase. We will mitigate the question mismatch issue by fine-tuning the question encoder in the concurrent learning phase.

The pretraining process of the retriever is described in Figure~\ref{fig:retriever}. Given a batch of $N$ question representations $V_q\in \mathbb{R}^{N \times 128}$ and their gold passage representations $V_p\in \mathbb{R}^{N\times 128}$, we obtain the retrieval scores for the batch by
\begin{equation}\label{eqn:batch-retriever-score}
\footnotesize
\begin{aligned}
\mathbf{S}_{rt}(V_q,  V_p) = V_q  V_p^\top
\end{aligned}
\end{equation}
where $\mathbf{S}_{rt}(V_q,  V_p) \in \mathbb{R}^{N\times N} $. The element $S_{i, j}$ in the $i$-th row and $j$-th column of $\mathbf{S}_{rt}(V_q,  V_p) $ represents $S_{rt}(q_i^{rw}, p_j)$. The objective is to maximize the probability of the gold passage for each question:
\begin{equation}\label{eqn:retriever-obj-pretrain}
\footnotesize
\begin{aligned}
P_{rt}(p_j|q_i^{rw}) = \frac{\exp{(S_{rt}(q_i^{rw}, p_j))}}{\sum_{j'=1}^N \exp{(S_{rt}(q_i^{rw}, p_{j'}))}}
\end{aligned}
\end{equation}
In other words, the passage set $\{p_j\}_{j=1}^N - \{p_i\}$ is considered as randomly sampled negative passages for $q_i$. The pretraining loss for this batch is then defined as follows.
\begin{equation}\label{eqn:pretrain-loss}
\footnotesize
\begin{aligned}
\mathcal{L}_{\text{pretrain}} = - \frac{1}{N}\sum_{i=1}^N\sum_{j=1}^N\mathds{1}\{j=i\} &\log{P_{rt}(p_j|q_i^{rw}) } 
\end{aligned}
\end{equation}
\citet{orqa} suggest that it is crucial to set the batch size $N$ to a large number because it makes the pretraining task more difficult and closer to what the retriever observes at test time. Therefore, we use two ALBERT models as the question encoder and the passage encoder. This doubles the batch size compared to that of using BERT models. The ALBERT models are fine-tuned.

We then encode all passages in the collection $C$ offline with the passage encoder and obtain a set of passage vectors.
Finally, we use Faiss\footnote{https://github.com/facebookresearch/faiss}, a library for efficient similarity search of dense vectors, to create an index for maximum inner product search. Retrieval is performed on a GPU during concurrent learning for faster training.

\subsubsection{\textbf{Concurrent Learning of the Retriever, Reranker, and Reader}}
As indicated in Figure~\ref{fig:model}, given the current question $q_k$, the history questions $\{q_i\}_{i=1}^{k-1}$, and the history window size $w$, we obtain the reformatted question for the retriever $q_k^{rt}$ and the reader $q_k^{rd}$. We first obtain the question representation of $q_k^{rt}$ using the question encoder in Equation~\ref{eqn:question-encoder}. We then retrieve the top $K_{rd}$ passages for the reader from the passage collection using the index we created offline. This set of top passages is denoted as $TK_{rd}$. The number of negative samples for retriever is limited by the CUDA memory in the retriever pretraining phase. In the concurrent learning phase, we can use a relatively large amount of negative samples to fine-tune the retriever at a low cost since all passages have been encoded offline. Therefore, we also retrieve the top $K_{rt}$ passages, where $K_{rt} > K_{rd}$, for an aggressive update of the retriever following \citet{orqa}.  This set of passages is denoted as $TK_{rt}$. If the gold passage of $q_k$ is not present in $TK_{rt}$ or $TK_{rd}$, we manually include it in the retrieval results. 
Formally, the retriever loss to fine-tune the question encoder in the retriever is defined as follows.
\begin{equation}\label{eqn:retriever-loss}
\footnotesize
\begin{aligned}
\mathcal{L}_{rt} = -\sum_{p_j \in TK_{rt}}\mathds{1}\{j=j_{rt}\} &\log{P_{rt}(p_j|q_k^{rt}) } 
\end{aligned}
\end{equation}
where $j_{rt}$ is the position of the gold passage in $TK_{rt}$.

Passages in $TK_{rd}$ are then fed into the reader/reranker module. This module conducts reading and reranking simultaneously. For every passage $p_j \in TK_{rd}$, we obtain a reranking score $S_{rr}(q_k^{rd}, p_j)$ following Equation~\ref{eqn:reranking-score}. We then compute the reranking probability and the reranking loss as follows.
\begin{equation}\label{eqn:rerank-prob}
\footnotesize
\begin{aligned}
P_{rr}(p_j|q_k^{rd}) = \frac{\exp{(S_{rr}(q_k^{rd}, p_j))}}{\sum_{p_{j'}\in TK_{rd}}\exp{(S_{rr}(q_k^{rd}, p_{j'}))}}
\end{aligned}
\end{equation}
\begin{equation}\label{eqn:rerank-loss}
\footnotesize
\begin{aligned}
\mathcal{L}_{rr} = -\sum_{p_j \in TK_{rd}}\mathds{1}\{j=j_{rd}\} &\log{P_{rr}(p_j|q_k^{rd}) } 
\end{aligned}
\end{equation}
where $j_{rd}$ is the position of the gold passage in $TK_{rd}$.

For the reader component, a standard BERT based machine comprehension model uses the cross entropy loss to maximize the probability of the true start and end tokens among all tokens in the given passage. Different from that, we apply the shared-normalization mechanism~\cite{shared-norm} to this step to maximize the probabilities of the true start and end tokens among all tokens from $TK_{rd}$. This makes the model produce start and end scores that are comparable across passages. The passages are encoded independently, and the shared-normalization is applied to all passages at the last step. For a passage $p_j \in TK_{rd} $, we obtain a start score $S_{s}(q_k^{rd}, p_j, \texttt{[m]})$ for every token \texttt{[m]} in the input sequence. The training loss for the start token is then defined as follows.
\begin{equation}\label{eqn:reader-start-prob}
\footnotesize
\begin{aligned}
P_{s}(q_k^{rd}, p_j, \texttt{[m]}) = \frac{\exp{(S_{s}(q_k^{rd}, p_j, \texttt{[m]}) )}}{\sum_{p_{j'} \in TK_{rd}}\sum_{\texttt{[m']}\in (q_k^{rd},\ p_{j'})} \exp{(S_{s}(q_k^{rd}, p_{j'}, \texttt{[m']}) )}}
\end{aligned}
\end{equation}
\begin{equation}\label{eqn:reader-start-loss}
\footnotesize
\begin{aligned}
\mathcal{L}_{s} = - \sum_{p_j \in TK_{rd}}\sum_{\texttt{[m]}\in (q_k^{rd},\ p_j)}\mathds{1}\{j=j_{rd}, \texttt{[m]} = \texttt{[S]}\} &\log{P_{s}(q_k^{rd}, p_j, \texttt{[m]})} 
\end{aligned}
\end{equation}
where \texttt{[S]} is the true start token in the gold passage. For unanswerable questions, we set the start and end tokens to \texttt{[CLS]}. The BERT encoder is fine-tuned.
The loss function of the end token $\mathcal{L}_{e}$ is defined in the same way. The reader loss is computed as follows.
\begin{equation}\label{eqn:reader-loss}
\footnotesize
\begin{aligned}
\mathcal{L}_{rd} = \frac{1}{2} (\mathcal{L}_{s} + \mathcal{L}_{e})
\end{aligned}
\end{equation}
Finally, the concurrent learning loss is computed as: 
\begin{equation}\label{eqn:loss}
\footnotesize
\begin{aligned}
\mathcal{L} = \mathcal{L}_{rt} + \mathcal{L}_{rr} + \mathcal{L}_{rd}
\end{aligned}
\end{equation}
Although the gradients of the reader/reranker do not back propagate to the retriever, we train these modules concurrently so that the reader/reranker can benefit from seeing more negative passages due to a dynamically changing set of retrieved passages $TK_{rd}$.
\subsection{Inference}
\label{subsec:model-inference}
Given the current question $q_k$, the history questions $\{q_i\}_{i=1}^{k-1}$, and the history window size $w$, we follow the same process in the concurrent learning phase to retrieve a set of relevant passages $TK_{rd}$. Note we do not manually include the gold passage in $TK_{rd}$ at inference time. For a passage $p_j \in TK_{rd} $, we obtain the retriever score $S_{rt}(q_k^{rt}, p_j)$ and the reranker score $S_{rr}(q_k^{rd}, p_j)$ following Equations \ref{eqn:retriever-score} and \ref{eqn:reranking-score}.
We then follow \citet{bert} to obtain the reader score using the start score $S_{s}(q_k^{rd}, p_j, \texttt{[m]}) $ and the end score $S_{s}(q_k^{rd}, p_j, \texttt{[m]}) $ in Equation~\ref{eqn:start-end-score} as follows.
\begin{equation}\label{eqn:reader-score}
\footnotesize
\begin{aligned}
&S_{rd}(q_k^{rd}, p_j, s) = \\
&\max_{[\texttt{m}_\texttt{s}], [\texttt{m}_\texttt{e}] \in (q_k^{rd},\ p_j)} S_{s}(q_k^{rd}, p_j, [\texttt{m}_\texttt{s}]) \ + 
 S_{e}(q_k^{rd}, p_j, [\texttt{m}_\texttt{e}]) 
\end{aligned}
\end{equation}
where $s$ is the answer span with the start token $[\texttt{m}_\texttt{s}]$ and end token $[\texttt{m}_\texttt{e}]$. To ensure tractability, we only consider the top 20 spans following convention~\cite{bert}.
Invalid predictions, including the cases where the start token comes after the end token, or the predicted span overlaps with the question part of the input sequence, are discarded. Finally, the overall score is defined as a function of the current question $q_k$, its history questions $\{q_i\}_{i=1}^{k-1}$, a history window size $w$, a retrieved passage $p_j$, and a answer span $s$ as in Figure~\ref{fig:model}:
\begin{equation}\label{eqn:total-score}
\footnotesize
\begin{aligned}
S(q_k, \{q_i\}_{i=1}^{k-1}, w, p_j, s) = & S_{rt}(q_k^{rt}, p_j) \ + \\
                                         & S_{rr}(q_k^{rd}, p_j) \ + \\
                                         & S_{rd}(q_k^{rd}, p_j, s) 
\end{aligned}
\end{equation}
The system outputs the answer span that has the largest overall score for each question in a conversation.

\section{Experimental Setups}
\label{sec:setup}
We now describe our experimental setups, including competing methods, evaluation metrics, and implementation details.

\subsection{\textbf{Competing Methods}}
\label{subsec:competing-methods}
To the best of our knowledge, there is no published work tackling the ORConvQA problem that we describe in Section~\ref{subsec:task}. There is, however, a rich body of work on single-turn open-domain QA, led by DrQA~\cite{drqa}. We can adapt such methods to a conversational setting by using the same history modeling method in our system. Given the effort to adapt such models to ORConvQA, we only compare to the original DrQA and the best model that we are aware of, BERTserini~\cite{bertserini}. 
To be specific, the competing methods are:
\begin{itemize}[leftmargin=1em]
    \item \textbf{DrQA}~\cite{drqa}. This model uses a TF-IDF retriever and an RNN based reader. We train this model on OR-QuAC dialogs with gold passages. At test time, the passages are retrieved with the retriever. This setting is consistent with DrQA's original setting.
    We do not use its distantly-supervised setting since we would like to adopt full supervision for all competing methods in this work.
    We start from their open-sourced implementation on GitHub.\footnote{https://github.com/facebookresearch/DrQA}
    \item \textbf{BERTserini}~\cite{bertserini}. This model uses a BM25 retriever from Anserini\footnote{http://anserini.io/} and a BERT reader. Their BERT reader is similar to ours, except that it does not support reranking and thus cannot benefit from multi-tasking learning. They study the granularity of retrieval, including article, paragraph, and sentence. They conclude that retrieval on a paragraph level gives the best overall performance. We only compare to the paragraph retrieval setting since it is the best and is consistent with our passage retrieval setting. We use the top 5 passages for the reader to be consistent with our setup. This baseline is our implementation since BERTserini's source code was not available at the time of our submission.
    \item \textbf{ORConvQA without history} (Ours w/o hist.). This is our model described in Section~\ref{sec:our-approach} with the history window size $w=0$. Note that the first question of a dialog is still included in the reformatted question for the retriever, as described in Section~\ref{subsec:model-retriever}. This model is our adaptation of the open-retrieval QA framework~\cite{orqa} to a conversational setting. We use a more direct and resource-efficient retriever pretraining method that is suitable for ConvQA. We also enable reranking in the reader component.
    \item \textbf{ORConvQA} (Ours). This is our full model described in Section~\ref{sec:our-approach}.
\end{itemize}

We adapt DrQA and BERTserini to a conversational setting using the same history modeling method in our model. It involves prepending history questions for reformatted questions for the retriever and the reader. For these models and our ORConvQA model, the history window size $w$ is tuned on the development set. We report their performance under the best history setting.

\subsection{\textbf{Evaluation Metrics}}
\label{subsec:metrics}
The word-level \textbf{F1} and the human equivalence score (\textbf{HEQ}) are two metrics provided by the QuAC challenge to evaluate ConvQA systems. F1 measures the overlap of the predicted answer span and the ground truth answer span. This is our most important metric since it evaluates the overall performance of the system.
HEQ computes the percentage of examples for which system F1 exceeds or matches human F1.
It measures whether a system can give answers as good as an average human. This metric is computed on a question level (HEQ-Q) and a dialog level (HEQ-D). 

In addition to F1 and HEQ, we also use the Mean Reciprocal Rank (\textbf{MRR}) and \textbf{Recall} to evaluate the retrieval performance for the retriever and reranker. The reciprocal rank of a query is the inverse of the rank of the first positive passage in the retrieved passages. MRR is the mean of the reciprocal ranks of all queries. This metric is computed for both the retriever and reranker. MRR is a reflection of how well these two components contribute to the overall score in Equation~\ref{eqn:total-score}. Recall is the fraction of the total amount of relevant passages that are retrieved. There is only one positive passage for each question in the training and development sets. In comparison, there could be more than one positive passage for a testing question since there are multiple reference answers per question provided by QuAC. Recall is computed for the retriever only since reranking does not impact this measure. This metric reflects whether the retriever can provide reasonable retrieval performance for the rest of the system. All retrieval metrics are computed for the top 5 passages that are retrieved for the reader/reranker.

\subsection{\textbf{Implementation Details}}
\label{subsec:details}
Our models are implemented with PyTorch\footnote{\url{https://pytorch.org/}} and the open-source implementation of ALBERT and BERT by Hugging Face.\footnote{\url{https://github.com/huggingface/transformers}} 
\subsubsection{\textbf{Retriever and Pretraining}}
We use two ALBERT Base (V1) models for the question and passage encoders. We set the max sequence length of the question encoder to 128, that of the passage encoder to 384, the training batch size to 16 per GPU, the number of training epochs to 12, and the learning rate to 5e-5. Models are trained with 4 NVIDIA TITAN X GPUs. 
We create a smaller collection to evaluate the retrieval performance by collecting the top 50 documents retrieved by TF-IDF for development questions. This allows us to do model selection in a scenario that is closer to how the retriever operates during concurrent learning.
We save checkpoints every 5,000 steps and evaluate on the development questions to select the best model for concurrent learning. The pretraining time for the retriever is 2.5 hours.

\subsubsection{\textbf{Reranker, Reader, and Concurrent Learning}}
We use the BERT Base (Uncased) model. We set the max sequence length to 512, the max question length to 125 (so that the passage length is at least 384 after accounting for a \texttt{[CLS]} and two \texttt{[SEP]} tokens), the training batch size to 2, the number of training epochs to 3, and the learning rate to 5e-5. We retrieve top 5 passages for the reader. We tune the number of passages to update retriever $K_{rt}$ and the history window size $w$ in Section~\ref{subsec:additional-analyses}. Models are trained with a NVIDIA TITAN X GPU. We take advantage of another TITAN X card for faster MIPS. All passage representations in our collection occupy 7.2 GB of CUDA memory. We save checkpoints every 5,000 steps and evaluate on the development set to select the best model for the test set. The time for concurrent learning is 20.0 hours.

For all model components, we use half precision for training as suggested in the Hugging Face repository to alleviate CUDA memory consumption. The warm up portion of the learning rate is 10\% of the total steps.

\section{Evaluation Results}
\label{sec:eval-results}
In this section, we present our evaluation results, ablation studies on system components, and more analyses on history window size and the number of passages to fine-tune the retriever.
\subsection{Main Evaluation Results}
\label{subsec:results}
\begin{table}[t]
\caption{Main evaluation results. ``Rt'' and ``Rr'' refers to ``Retriever'' and ``Reranker''. $\ddagger$ means statistically significant improvement over the strongest baseline with $p < 0.05$.
}
\label{tab:results}
\vspace{-0.3cm}
\begin{tabular}{@{}clllll@{}}
\toprule
\multicolumn{2}{l}{Settings}      & DrQA   & BERTserini & Ours w/o hist. & Ours            \\ \midrule
\multirow{6}{*}{Dev}  & F1        & 4.5    & 19.3       & 24.0         & \textbf{26.9}$^\ddagger$   \\
                      & HEQ-Q     & 0.0    & 14.1       & 15.2         & \textbf{17.5}   \\
                      & HEQ-D     & 0.0    & \textbf{0.2} & \textbf{0.2} & \textbf{0.2}    \\
                      & Rt MRR    & 0.1151 & 0.1767     & 0.4012       & \textbf{0.4286}$^\ddagger$ \\
                      & Rr MRR    & N/A    & N/A        & 0.4472       & \textbf{0.5209}$^\ddagger$ \\
                      & Rt Recall & 0.2000 & 0.2656     & 0.5271       & \textbf{0.5714}$^\ddagger$ \\ \midrule
\multirow{6}{*}{Test} & F1        & 6.3    & 26.0       & 26.3         & \textbf{29.4}$^\ddagger$   \\
                      & HEQ-Q     & 0.1    & 20.4       & 20.7         & \textbf{24.1}   \\
                      & HEQ-D     & 0.0    & 0.1        & 0.4          & \textbf{0.6}    \\
                      & Rt MRR    & 0.1574 & 0.1784     & 0.1979       & \textbf{0.2246}$^\ddagger$ \\
                      & Rr MRR    & N/A    & N/A        & 0.2702       & \textbf{0.3127}$^\ddagger$ \\
                      & Rt Recall & 0.2253 & 0.2507     & 0.2859       & \textbf{0.3141}$^\ddagger$ \\ \bottomrule
\end{tabular}

\end{table}
We report the main evaluation results in Table~\ref{tab:results}. We tune the history window size $w$ for all models that consider history and report their performances under the best history setting. The best history settings for DrQA, BERTserini, and Ours are $w=$5, 2, and 6 respectively. We summarize our observations as follows:
\begin{enumerate}[leftmargin=1.5em]
    \item We observe that DrQA has poor performance. The main reason for this lies in the reader component. The RNN based reader in DrQA cannot produce representations that are as good as the readers based on a pretrained BERT in the rest of the competing models. More importantly, the DrQA reader cannot handle unanswerable questions natively.
    \item BERTserini has a significant improvement over DrQA and serves as a much stronger baseline. It addresses the issues in DrQA by using a BERT reader that can handle unanswerable questions. BM25 in Anserini also gives better retrieval performance.
    \item Our model without any history manages to perform on par with BERTserini that considers history on the test set. In particular, our learned retriever achieves higher performance on retrieval metrics. Since our reader is similar to that of BERTserini, the overall performance gain mostly comes from our learned retriever. This verifies the observation in \citet{orqa} in a conversational setting that a learned retriever is crucial if the information-seeker is genuinely seeking an answer. The margins are substantially larger on the development set, presumably because the best pretrained retriever model is selected based on the development performance.
    \item Our model with history obtains statistically significant improvement over the strongest baseline with $p < 0.05$ tested by the Student's paired t-test. This demonstrates the effectiveness of our model. This also indicates that incorporating conversation history is essential for ORConvQA, as expected. More analyses on the history window size are presented in Section\ref{subsubsec:additional-analyses-history}. In addition, we observe that the reranker consistently outperforms the retriever. This suggests that although reranking is more expensive as it jointly models the question and the passage, it enjoys better performance than the retriever that models the question and the passage separately.

\end{enumerate}

\subsection{Ablation Studies}
\label{subsec:ablation}
\begin{table}[t]
\caption{Results of ablation studies. ``Rt'' and ``Rr'' refers to ``Retriever'' and ``Reranker'' respectively. $\ddagger$ and $\dagger$ means statistically significant performance \textit{decrease} compared to the full system with $p < 0.05$ and $p<0.1$ respectively. 
}
\label{tab:ablation}
\vspace{-0.3cm}
\begin{tabular}{@{}clllll@{}}
\toprule
\multicolumn{2}{l}{Settings}      & \begin{tabular}[c]{@{}l@{}}Full \\ system\end{tabular}            & \begin{tabular}[c]{@{}l@{}}w/o \\ rerank\end{tabular} & \begin{tabular}[c]{@{}l@{}}w/o learned \\ retriever\end{tabular} & \begin{tabular}[c]{@{}l@{}}w/o first q \\ for retriever\end{tabular} \\ \midrule
\multirow{6}{*}{Dev}  & F1        & \textbf{26.9}   & 25.9$^\dagger$                                                  & 17.1$^\ddagger$                                                             & 24.6$^\ddagger$                                                                 \\
                      & HEQ-Q     & \textbf{17.5}   & 16.8                                                  & 11.1                                                             & 15.5                                                                 \\
                      & HEQ-D     & 0.2             & 0.2                                                   & 0.0                                                              & \textbf{0.4}                                                         \\
                      & Rt MRR    & \textbf{0.4286} & 0.4031$^\ddagger$                                                & 0.1162$^\ddagger$                                                           & 0.3937$^\ddagger$                                                               \\
                      & Rr MRR    & \textbf{0.5209} & N/A                                                    & 0.1895$^\ddagger$                                                           & 0.4674$^\ddagger$                                                               \\
                      & Rt Recall & \textbf{0.5714} & 0.5411$^\ddagger$                                                & 0.2032$^\ddagger$                                                           & 0.5122$^\ddagger$                                                               \\ \midrule
\multirow{6}{*}{Test} & F1        & \textbf{29.4}   & 27.7$^\ddagger$                                                  & 24.7$^\ddagger$                                                             & 27.1$^\ddagger$                                                                 \\
                      & HEQ-Q     & \textbf{24.1}   & 22.2                                                  & 18.1                                                             & 21.3                                                                 \\
                      & HEQ-D     & 0.6             & 0.9                                                   & 0.5                                                              & \textbf{1.0}                                                         \\
                      & Rt MRR    & \textbf{0.2246} & 0.2166$^\ddagger$                                                & 0.1603$^\ddagger$                                                           & 0.2092$^\ddagger$                                                               \\
                      & Rr MRR    & \textbf{0.3127} & N/A                                                    & 0.2130$^\ddagger$                                                           & 0.2870$^\ddagger$                                                               \\
                      & Rt Recall & \textbf{0.3141} & 0.3059$^\dagger$                                                & 0.2270$^\ddagger$                                                           & 0.2918$^\ddagger$                                                               \\ \bottomrule
\end{tabular}

\end{table}

Section~\ref{subsec:results} has shown the effectiveness of our model. This model performance is closely related to several design choices we made. In this section, we conduct ablation studies on our best model in Table~\ref{tab:results} to investigate the contributions of each design choice. Specifically, we have three ablation settings as follows.
\begin{itemize}[leftmargin=1em]
    \item \textbf{ORConvQA w/o reranker}. We introduce reranking to the system as one of the differences from previous works~\cite{orqa,Das2019MultistepRI}. In this ablation setting, we remove the reranking loss in Equation~\ref{eqn:loss} so that the encoder in the reader is not fine-tuned by the reranking objective. Naturally, we also do not use the reranking score in the overall score in Equation~\ref{eqn:total-score}.
    \item \textbf{ORConvQA w/o learned retriever}. We replace our learned retriever with DrQA's TF-IDF retriever.
    \item \textbf{ORConvQA w/o first question (q) for retriever}. We do not manually include the first question of a dialog in the reformatted question for the retriever.
\end{itemize}

The ablation results are presented in Table~\ref{tab:ablation}. The following are our observations.
\begin{enumerate}[leftmargin=1.5em]
    \item By removing the reranker from the full system, we observe a degradation in the overall performance. Although the reranking loss does not influence the retriever, the retriever performance also decreases. This is because that the ablated system gives the best development performance earlier than the full system during training. The reason behind this is that the reader overfits before the retriever has enough fine-tuning to produce reasonable retrieval performance. This verifies our assumption that the encoder in the reader/reranker benefits from a regularization effect by optimizing for the additional reranking task.
    \item Replacing the learned retriever with TF-IDF causes a dramatic performance drop. This further verifies our observation in Section~\ref{subsec:results} that a learned retriever is crucial for ORConvQA.
    \item When we do not additionally include the first question of the dialog in the reformatted question for the retriever, we observe a statistically significant performance decrease on most of the metrics. 
    This validates our observation during data construction that the initial question of a dialog often contains a general information need that is pertinent to the entire dialog. 
    By including the initial questions, the retriever can retrieve passages that are more relevant to the information need. 
    The performance drop is less substantial than we anticipated. This is probably because the history window size of 6 has already covered the initial question for more than half of the questions, given that the number of history turns per question has a median of 3.
\end{enumerate}

\subsection{Additional Analyses}
\label{subsec:additional-analyses}

\subsubsection{\textbf{Impact of history window size}}
\label{subsubsec:additional-analyses-history}
\begin{figure}[t]
    \centering
    \begin{subfigure}[b]{0.23\textwidth}
        \includegraphics[width=\textwidth]{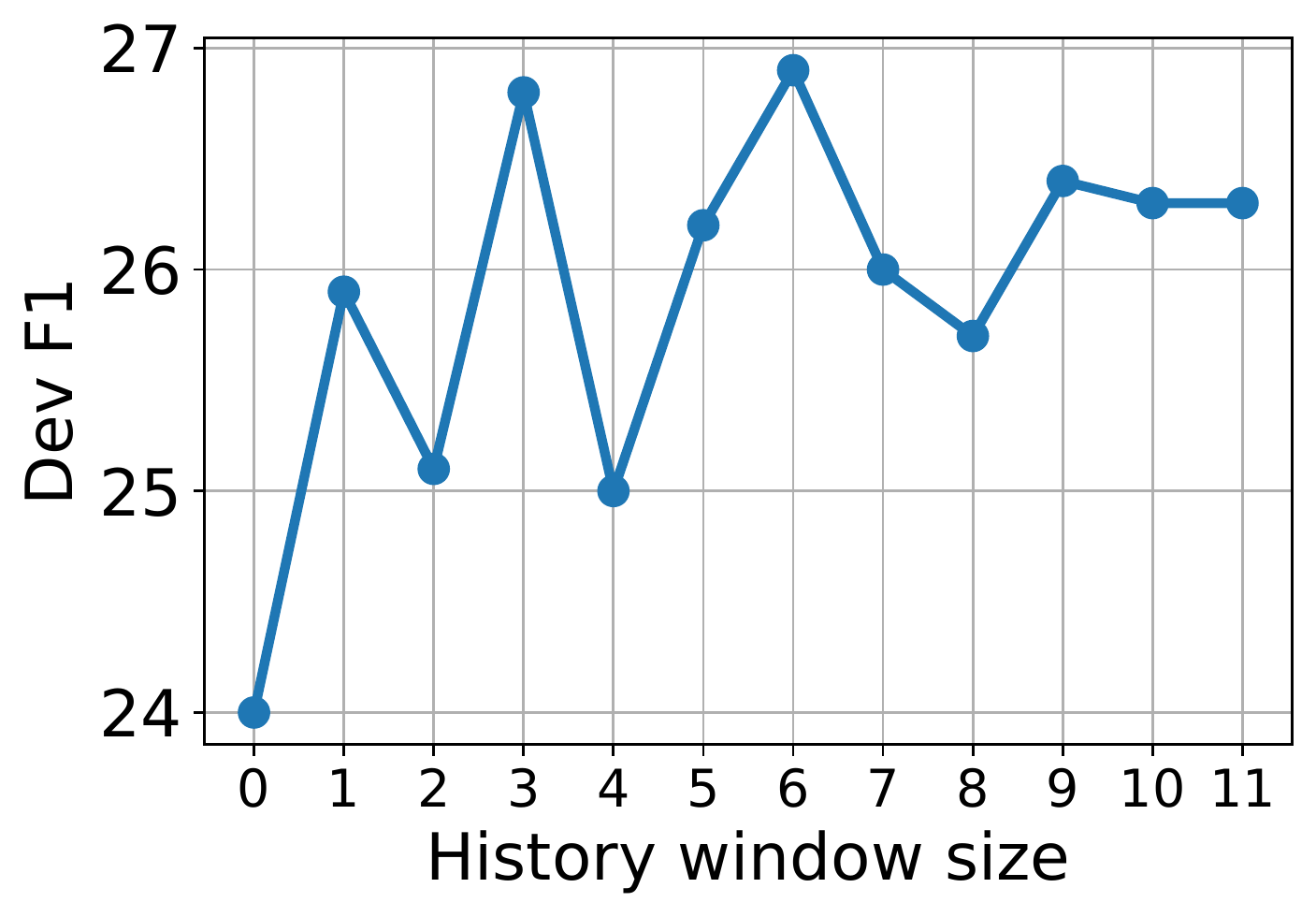}
        \vspace{-0.3cm}
        \caption{Dev overall performance.}
        \label{fig:history_f1}
    \end{subfigure}
    \begin{subfigure}[b]{0.23\textwidth}
        \includegraphics[width=\textwidth]{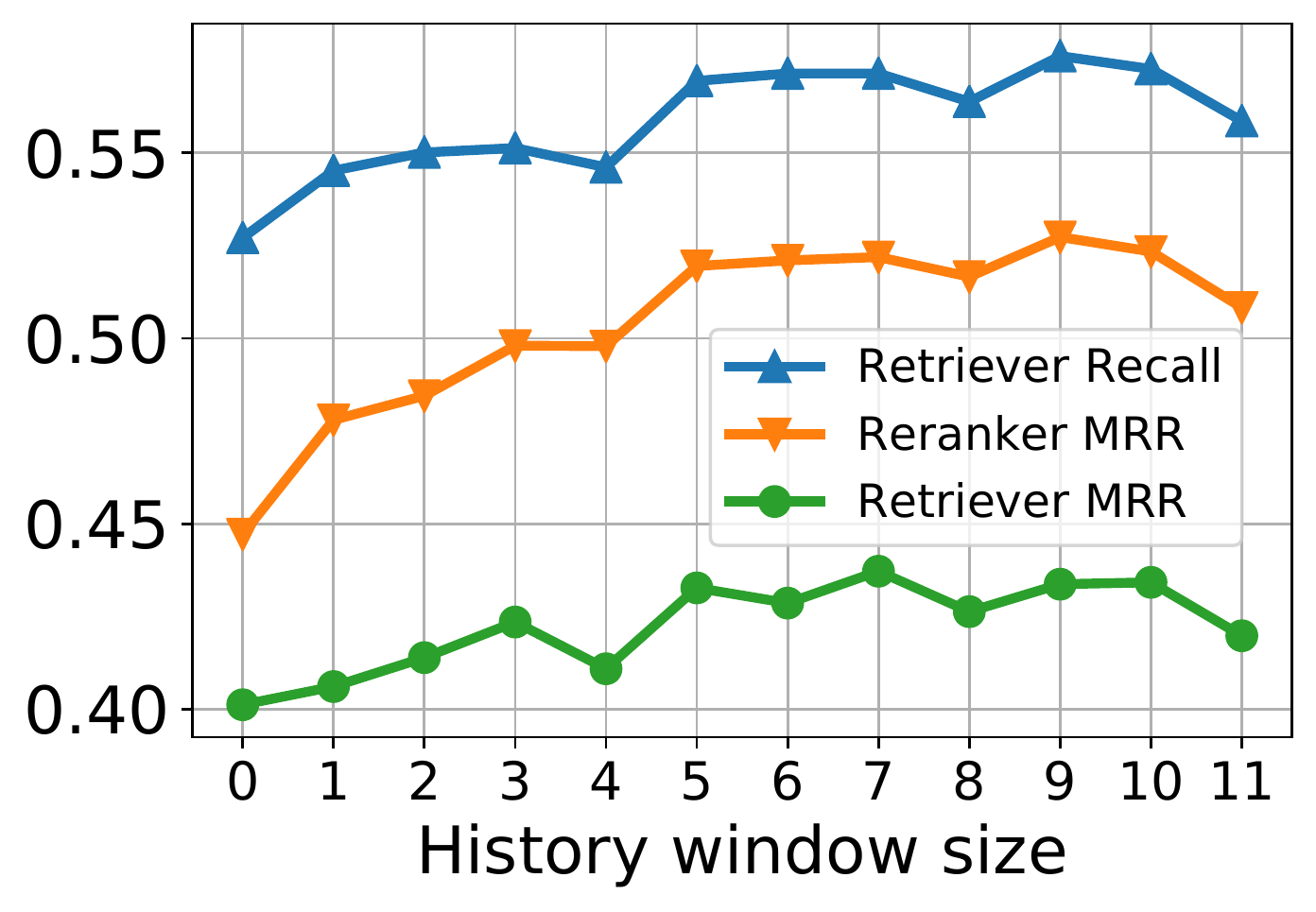}
        \vspace{-0.3cm}
        \caption{Dev retrieval performance.}
        \label{fig:history_retriever}
    \end{subfigure}
    \vspace{-0.3cm}
    \caption{Impact of history window size $w$.}
    \label{fig:history}
\end{figure}

Leveraging conversation history is an integral part of a ConvQA system and has not been well studied in an open-retrieval setting. In this section, we study the impact of the history window size $w$ on the system performance. The results are presented in Figure~\ref{fig:history}. 

In Figure~\ref{fig:history_f1}, we observe that incorporating any number of history turns outperforms no history at all. Although fluctuating, the overall performance first increases then decreases, with the peak value at $w=6$. In Figure~\ref{fig:history_retriever}, we observe that all retrieval metrics generally grow as we incorporate more conversation history. This suggests that the additional history turns we prepend are useful for matching and retrieval in most cases. Since we have reserved 125 tokens for the reformatted question in the BERT input sequence as reported in Section~\ref{subsec:details}, we show less degradation in the performance than previous work~\cite{hae} when we prepend more history.

It is intriguing that the retriever recall, the most important retrieval metric, shows a trimodal distribution. This could be due to the ``topic return'' phenomenon mentioned in \citet{Yatskar2018AQC}. Given the current question in a dialog, an adjacent turn is typically more useful than a distant turn to reveal the information need of the current turn. In other words, the utility of a history turn decreases as the distance between itself and the current turn increases. This utility trend shifts when the current turn is returning to the topic that has been discussed in a distant history turn. The trimodal distribution could imply that a topic return phenomenon typically happens five turns or nine turns away from the current turn. Moreover, the valley values of the trimodal distribution of retriever recall are consistent with those of the F1 curve in Figure~\ref{fig:history_f1}, suggesting that the fluctuation in the overall performance can be explained by the variation in retriever performance.

\subsubsection{\textbf{Impact of the number of passages to update retriever}}
\label{subsubsec:additional-analyses-k}
\begin{figure}[t]
    \centering
    \begin{subfigure}[b]{0.23\textwidth}
        \includegraphics[width=\textwidth]{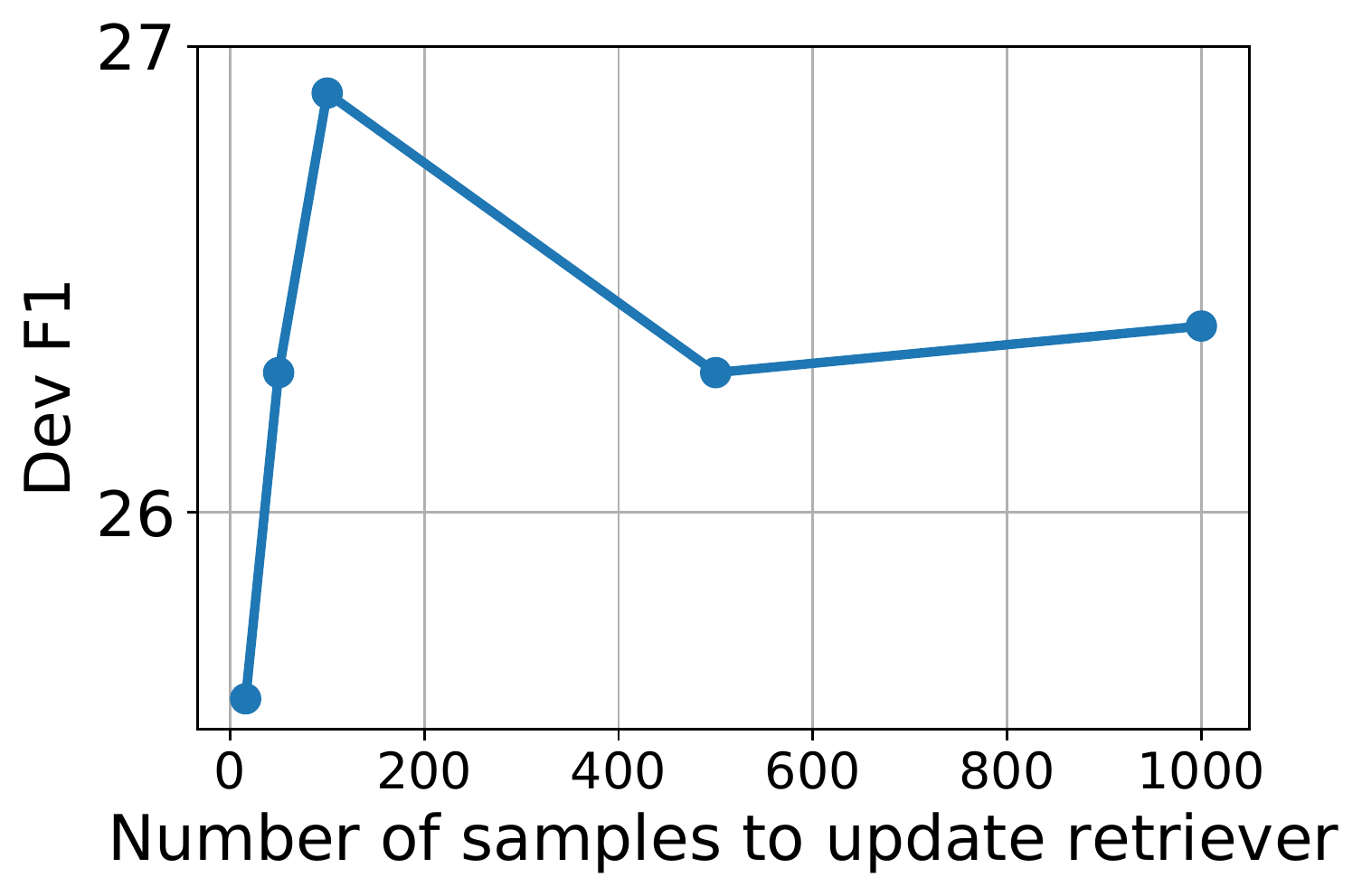}
        \vspace{-0.2cm}
        \caption{Dev overall performance.}
        \label{fig:krt_f1}
    \end{subfigure}
    \begin{subfigure}[b]{0.23\textwidth}
        \includegraphics[width=\textwidth]{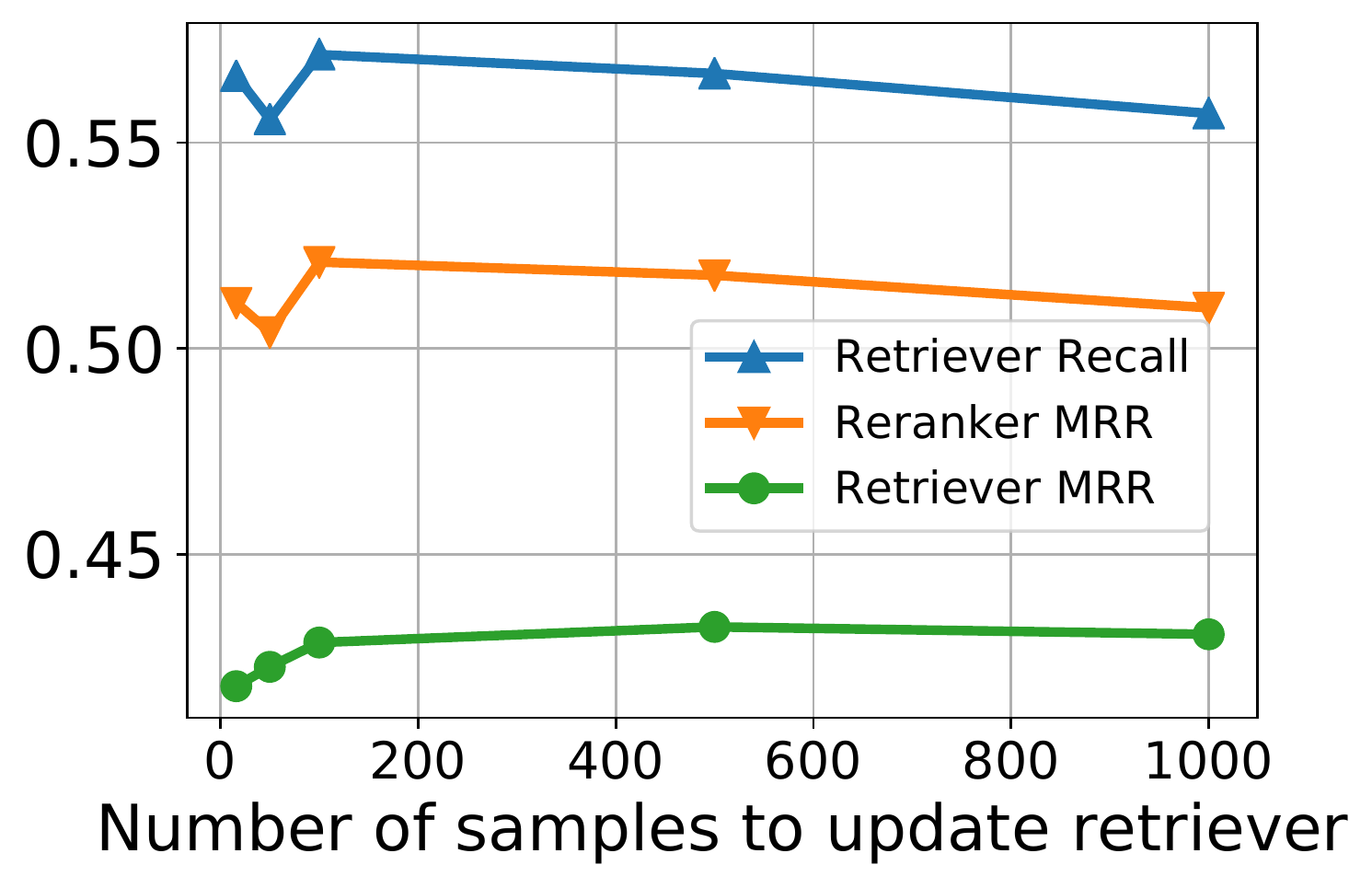}
        \vspace{-0.2cm}
        \caption{Dev retrieval performance.}
        \label{fig:krt_retriever}
    \end{subfigure}
    \vspace{-0.2cm}
    \caption{Impact of \# samples to update retriever $K_{rt}$.}
    \label{fig:krt}
\end{figure}

\citet{orqa} suggest that it is crucial to set the batch size $N$ in the retriever pretraining phase as large as possible because it makes the pretraining task more difficult and closer to what the retriever observes at test time. During pretraining, we set $N$ to 16 as reported in Section~\ref{subsec:details}, meaning that we have 16 passages per question to train the retriever. At the concurrent learning phase, we can increase this number to fine-tune the question encoder in the retriever at a low cost since all passages have been encoded offline. Therefore, we investigate how helpful it is to increase the number of passages $K_{rt}$ to fine-tune the retriever during concurrent learning. The choices of $K_{rt}$ are [16, 50, 100, 500, 1000]. We sample the choices of $K_{rt}$ unevenly and with large gaps so that the trends are clear. The results are presented in Figure~\ref{fig:krt}.

We observe that $K_{rt} = 100$ gives the best overall performance and retriever recall. Using a smaller and larger number both give a sub-optimal performance. Although a smaller value is closer to what we use for pretraining, the retriever cannot aggressively learn from enough negative passages. On the contrary, if we use a $K_{rt}$ value that is progressively larger than that of the pretraining time, the mismatch of supervision signals also leads to inferior performance.
\section{Conclusions and Future Work} 
\label{sec:conclusion}
In this work, we introduce an open-retrieval conversational QA setting as a further step towards conversational search. We create a dataset, OR-QuAC, by aggregating existing data to facilitate research on ORConvQA. We build an end-to-end system for ORConvQA, featuring a retriever, a reranker, and a reader that are all based on Transformers. Our extensive experiments on OR-QuAC demonstrate that a learnable retriever is crucial in the ORConvQA setting. We further show that our system can make a substantial improvement when we enable history modeling in all system components. Moreover, we show that the additional reranker component contributes to the model performance by providing a regularization effect. Finally, we demonstrate that the initial question of each dialog is essential for our system to understand the user's information need. For future work, we would like to address the limitations of this work by studying weak supervision methods for information-seeking conversations and a retriever that is not only learnable but also tunable by the downstream task. In addition, we will investigate more effective history modeling methods.

\begin{acks}
This work was supported in part by the Center for Intelligent Information Retrieval, in part by NSF IIS-1715095, and in part by China Postdoctoral Science Foundation (No. 2019M652038). Any opinions, findings and conclusions or recommendations expressed in this material are those of the authors and do not necessarily reflect those of the sponsor.
\end{acks}

\bibliographystyle{abbrvnat}
\balance 
\bibliography{acmart} 

\end{document}